\documentclass[aps,prd,twocolumn]{revtex4-1}

\usepackage{graphicx}  
\graphicspath{ {images/} } 
\usepackage{latexsym}
\usepackage{slashed}
\usepackage{dcolumn}   
\usepackage{bm}        
\usepackage{amsmath}
\usepackage{amssymb}   
\usepackage{longtable}
\usepackage{float}
\newcommand{\be}{\begin{equation}}
\newcommand{\ee}{\end{equation}}
\newcommand{\bea}{\begin{eqnarray}}
\newcommand{\eea}{\end{eqnarray}}
\newcommand{\bma}{\begin{matrix}}
\newcommand{\ema}{\end{matrix}}

\newcommand{\bml}{\begin{mathletters}}
\newcommand{\eml}{\end{mathletters}}
\newcommand{\bes}{\begin{subequations}} 
\newcommand{\ees}{\end{subequations}}

\newcommand{\bi}{\begin{itemize}}
\newcommand{\ei}{\end{itemize}}
\newcommand{\gev}{~{\rm GeV}}

\begin{document}
\title{Constraints from the S-parameter on models with scalar triplets}
\author{P. Q. Hung}
\affiliation{Department of Physics, University of Virginia,
Charlottesville, VA 22904-4714, USA}

\date{\today}

\begin{abstract}
Models with scalar triplets can have a large impact on the electroweak precision S-parameter and have to be dealt "with extreme prejudice". Such models, and in particular the Georgi-Machacek (GM) model, involve a complex triplet scalar containing doubly-charged Higgses plus a real triplet and are of recent phenomenological interests. It turns out that large mass splittings within the triplet will not be allowed by the S-parameter in the absence of extra fermionic degrees of freedom. Simply speaking, having only an extra triplet of scalars severely restricts the mass splitting within that triplet even with the present precision on the S-parameter which will further be improved with the proposed Higgs factories.
\end{abstract}

\maketitle

\section{Introduction}
The present dearth of new physics beyond the Standard Model (BSM) has motivated the community to push for wider searches for new particles which may or may not show up at the LHC. It goes without saying that the minimal SM is highly unsatisfactory, despite of its successes in describing the known interactions except for gravity. For example, the nature of the spontaneous symmetry breaking of the SM is up-in-the-air since the only knowledge that we have about the 125-GeV $0^+$ particle discovered in 2012 (without making extra assumptions) was its signal strength (production cross section x branching ratio, normalized by the corresponding quantity evaluated in the SM). It is not unreasonable to ask if other types of Higgs scalars exist and which can contribute to the spontaneous breakdown of the SM. 

Of particular interest are Georgi-Machacek (GM)-type of models \cite{GM} which contain a complex triplet $\tilde{\chi}=(\tilde{\chi}^{++}, \tilde{\chi}^+, \tilde{\chi}^0)$ and a real triplet $\xi=(\xi^+, \xi^0, \xi^-)$. As explained in \cite{}, these two triplets are required to coexist and to have equal vacuum expectation values (VEV) in order to maintain a custodial symmetry at tree level so that $\rho=m_{W}^2/m_{Z}^2 \cos^2 \theta_W =1$. Without any doubt, the discovery of these triplet Higgs scalars would certainly be very exciting for, not only they might point to the origin of neutrino masses \cite{pqnur}, they can also elucidate the mechanism for the spontaneous breakdown of the SM. However, a mere introduction of the GM triplets into the SM comes at a cost: the electroweak precision parameter S severely limits its parameter space. This fact which we will show below restraints in a significant way the phenomenology of the GM triplets.


On the other hand, there are models which contain, in the $SU(2)_L$ sector, one scalar triplet alone in addition to the usual SM particle content \cite{models}. As we shall see below, the contribution to the S-parameter is generally negative and can be large in magnitude  for a wide range of parameters (e.g. triplet mass splitting). However, the present constraint on S severely restricts the mass parameter space and will be even more so with better determinations of S from proposed $e^{+}e^{-}$ colliders.

\section{Contributions to the S-parameter from $SU(2)_L$ scalar triplets}

In this section, we will study the implications the S-parameter on models containing $SU(2)$ scalar triplets, starting with the present value of S to its possible determination at future $e^{+}e^{-}$ colliders. From hereon, we shall denote the {\em new physics contribution} to S by $S_{new}$. We first start with the ideal case where $S_{new}=0$ (or very small) which coincides with that of the Standard Model and ask under which conditions the new physics, if any, is indistinguishable from the SM as far as $S_{new}$ is concerned. We then move on to discuss the more realistic cases: the present constraint $S_{new}=0.02 \pm 0.07$ followed by expectations for $S_{new}$ from data obtained at the proposed $e^{+}e^{-}$ colliders.

Let us suppose that $S_{new}=0$. We propose the following statement:
{\bf {\em $S_{new}=0$ is consistent with the minimal Standard Model and with models in which this occurs because of cancellations among sectors that can contribute to $S_{new}$. Extensions of the SM with only $SU(2)_L$ Higgs triplets and nothing else extra will be consistent with $S_{new}=0$ only in the extreme fine tuning situation.}} 

We first start with a simplified discussion with just one $SU(2)_L$ scalar triplet (typical of Left-Right (LR) and Type II Seesaw models) and move on to a more involved triplet structure such as the GM model and also the Electroweak-scale right-handed neutrino (EW-$\nu_R$) model \cite{pqnur}. Notice that, in the former class of models, it is well-known that the VEV of the triplet scalar has to be small in order to accommodate $\rho \approx 1$ at tree level while the GM-class of models can accommodate VEVs of the order of the electroweak scale $\Lambda_{EW} \sim 246 \gev$. 

The contribution to the S-parameter from an $SU(2)_L$ scalar triplet has been computed in \cite{mehrdad} and later recalculated within the framework of the Electroweak-scale right-handed neutrino model (to be referred to as the EW-$\nu_R$  model from hereon) \cite{ajinkya}. It takes the form \cite{mehrdad}
\be
\label{S}
S_{triplet} =\frac{2}{9 \pi}\{\frac{1}{3}\ln \zeta + 8 \int_{0}^{1} dx x(1-x) \ln (x+ \zeta(1-x)) \},
\ee
where, in \cite{mehrdad}, we have defined $\zeta=(1-2\beta^2)/(1+\beta^2))$ with $\beta=m'/m$ and $m'$ is the mass splitting parameter. The degenerate case corresponds to $m'=0$ or $\beta=0$. Below is the plot of S ($\equiv S_{triplet}$) versus $\beta$.
\begin{figure}[H]
\centering
    \includegraphics[scale=0.62]{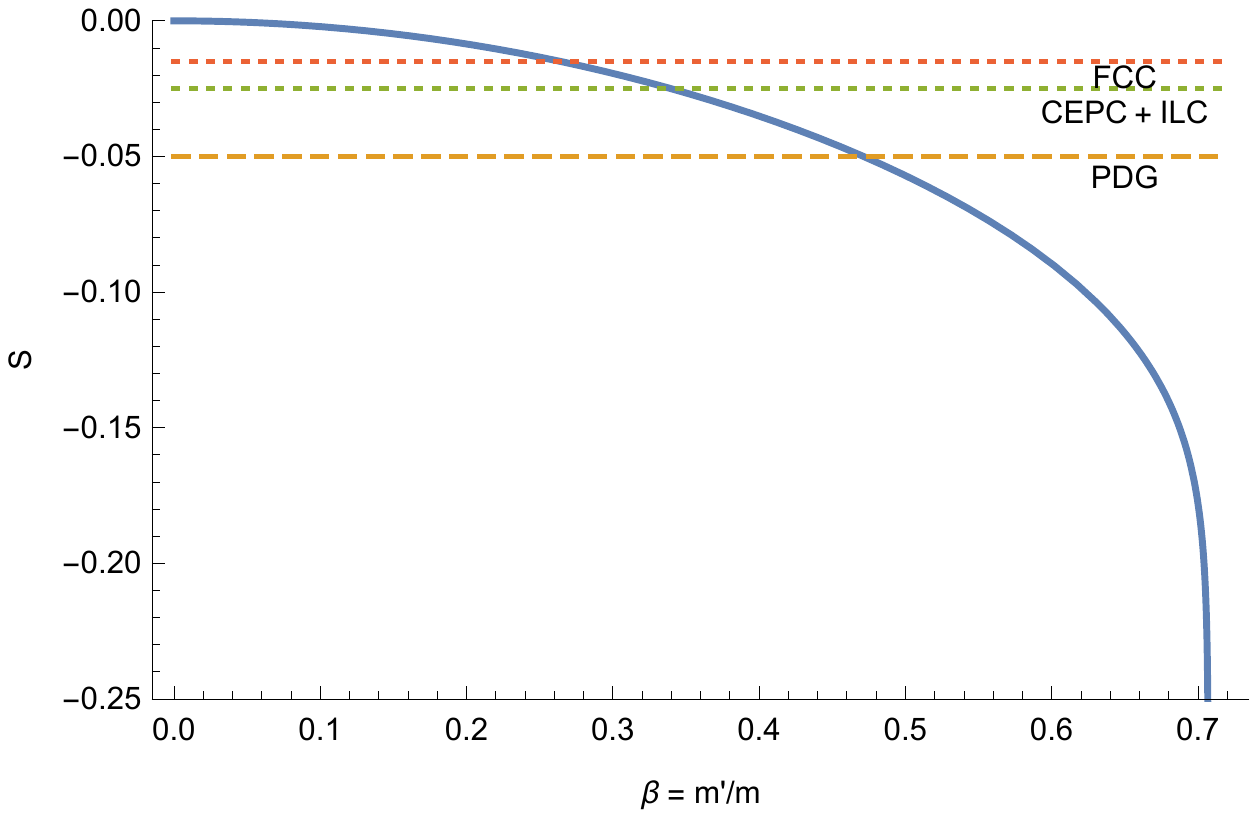} 
\caption{{\small S vs the mass splitting ratio $\beta=\frac{m'}{m}$. The dashed and the dotted lines represent the current precision (PDG) and the projected precision for the ILC and CEPC colliders.}}
\label{S1}
\end{figure}
 As one can see from Fig.~\ref{S1}, the S-parameter can be very negative for a large mass splitting. Also shown are the current precision $S=-0.05$ ($S=0.02 \pm 0.07$ from PDG) \cite{PDG} and the projected precision from the planned ILC,CEPC, and FCC$_{ee}$ \cite{reece}. This simple model shows that the present constraint from S limits the mass splitting within the triplet to be $m^{'} /m < 0.48$ due to the fact that the error on S is still large. We can expect that, with a better precision that can be obtained at planned Higgs factories such as the ILC and CEPC, the mass splitting will be narrowed down to $\sim 0.35$ while at the planned FCC$_{ee}$, it is $\sim 0.25$ (a number which will be updated in the near future). 
 
Fig.~\ref{S1} shows that $S_{new}=0$ only for the {\em degenerate case} $m^{'}=0$. This, of course, is a highly fine-tuned situation. As we show below, the case $S_{new}\approx 0$ can be realized for large triplet mass splittings if there exists mirror fermions whose contributions to $S$ can cancel those from the triplets with appropriate choices of parameters. We would like to stress that there are {\em several of such choices}.
  
 Phenomenological analyses of models containing, beside SM particles, one Higgs triplet of $SU(2)_L$ will have to seriously take into account constraints coming from the S-parameter which limits the mass parameter space of members of that triplet. These constraints are shown in Fig.~\ref{S1}. If we denote the triplet by $(\Delta^{0}, \Delta^{+}, \Delta^{++})$ and if, for example, $M_{\Delta^{+}} \sim 200 \gev$, then $M_{\Delta^{++}} < 300 \gev$ from the present constraint $m^{'} /m < 0.48$. On the other hand, present LHC
 limits on the mass of the doubly-charged scalar $\Delta^{++})$ based on rather special assumptions put  $M_{\Delta^{++}} > 770-870 \gev$ for $B( \Delta^{++} \rightarrow l^{+} l^{+})=100\%$ and $M_{\Delta^{++}} > 450 \gev$ for $B( \Delta^{++} \rightarrow l^{+} l^{+})=10\%$ \cite{atlas}, and $M_{\Delta^{++}} > 800-820 \gev$ for $B( \Delta^{++} \rightarrow l^{+} l^{+})=100\%$ ($e^{+} e^{+},...$) and $M_{\Delta^{++}} > 643-714 \gev$ for $B( \Delta^{++} \rightarrow l^{+} l^{',+})=100\%$ ($e^{+} \mu^{+},...$) \cite{cms}. The aforementioned S-parameter constraint would then imply that $M_{\Delta^{+}} > 234-452 \gev$ (with branching ratios for $10\%$ to $100\%$) for the ATLAS bounds and $M_{\Delta^{+}} > 334-426 \gev$ for CMS. (One should of course take these bounds with a grain of salt.) 
The GM model contains a complex triplet $\tilde{\chi} (Y/2=1)=(\chi^{++}, \chi^{+}, \chi^{0})$ and a real triplet $\xi \ (Y/2 = 0)= (\xi^+, \xi^0, \xi^-)$ where the weak hypercharge is $Y/2$. These scalars can be put in a $3 \times 3$ matrix, a $(3,3)$ representation of a global $SU(2)_L \times SU(2)_R$ symmetry. 
As emphasized in \cite{GM}, $\langle \chi^0 \rangle = \langle \xi^0 \rangle=v_M$ gives $SU(2)_L \times SU(2)_R \rightarrow SU(2)_D$ where $SU(2)_D$ is the custodial symmetry which preserves, at tree level, $\rho=1$ or $M_{W}^2 = M_{Z}^2 \cos^{2}\theta_W$. As a result, the GM model allows $v_M =O(\Lambda_{EW} \approx 246 \gev)$. This feature was used by \cite{pqnur} in the construction of the EW-$\nu_R$ model in which non-sterile $\nu_R$'s have masses proportional to $\Lambda_{EW} \approx 246 \gev$ and can be searched for at the LHC. 

Including the usual Higgs doublet, the following physical scalars are classified as representations of the custodial symmetry $SU(2)_D$.
		$\text{Five-plet (quintet)}:  H_5^{\pm\pm},\; H_5^\pm,\; H_5^0$;
		$\text{Triplet}: H_{3}^\pm,\; H_{3}^0$;
		$\text{Two singlets}: H_1^0,\; H_1^{0\prime}$.
 The contribution of these physical scalar states to the S-parameter has been computed in \cite{ajinkya} for the Electroweak-scale right-handed neutrino (EW-$\nu_R$) model \cite{pqnur}. The model contains mirror fermions and the GM triplets. The calculations were done separately for the scalar and mirror fermion sectors. In \cite{ajinkya}, we have carried out computations for the S and T parameters by varying the masses of the various scalars as well as the mixing parameter $\sin \theta_H$. We also varied masses of the mirror fermions. 
 Details of the calculations can be found in \cite{ajinkya}. In what follows, $\tilde{S}$ and $\tilde{T}$ refer to new physics contributions to the S and T parameters.
\setlength{\parskip}{3mm}

\begin{figure}[t]
\centering
    \includegraphics[scale=0.3]{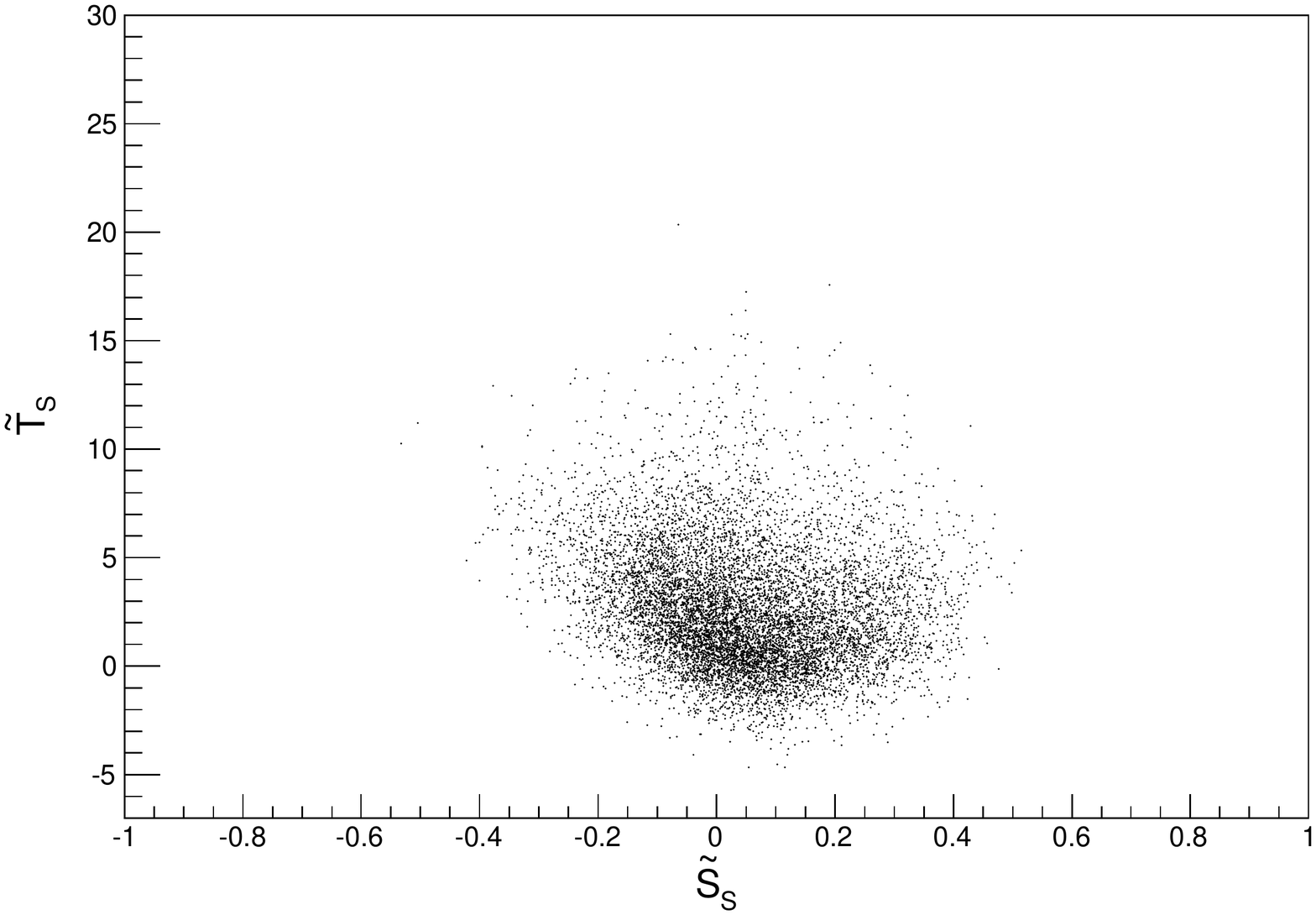} 
\vspace{-2pt}
    \includegraphics[scale=0.3]{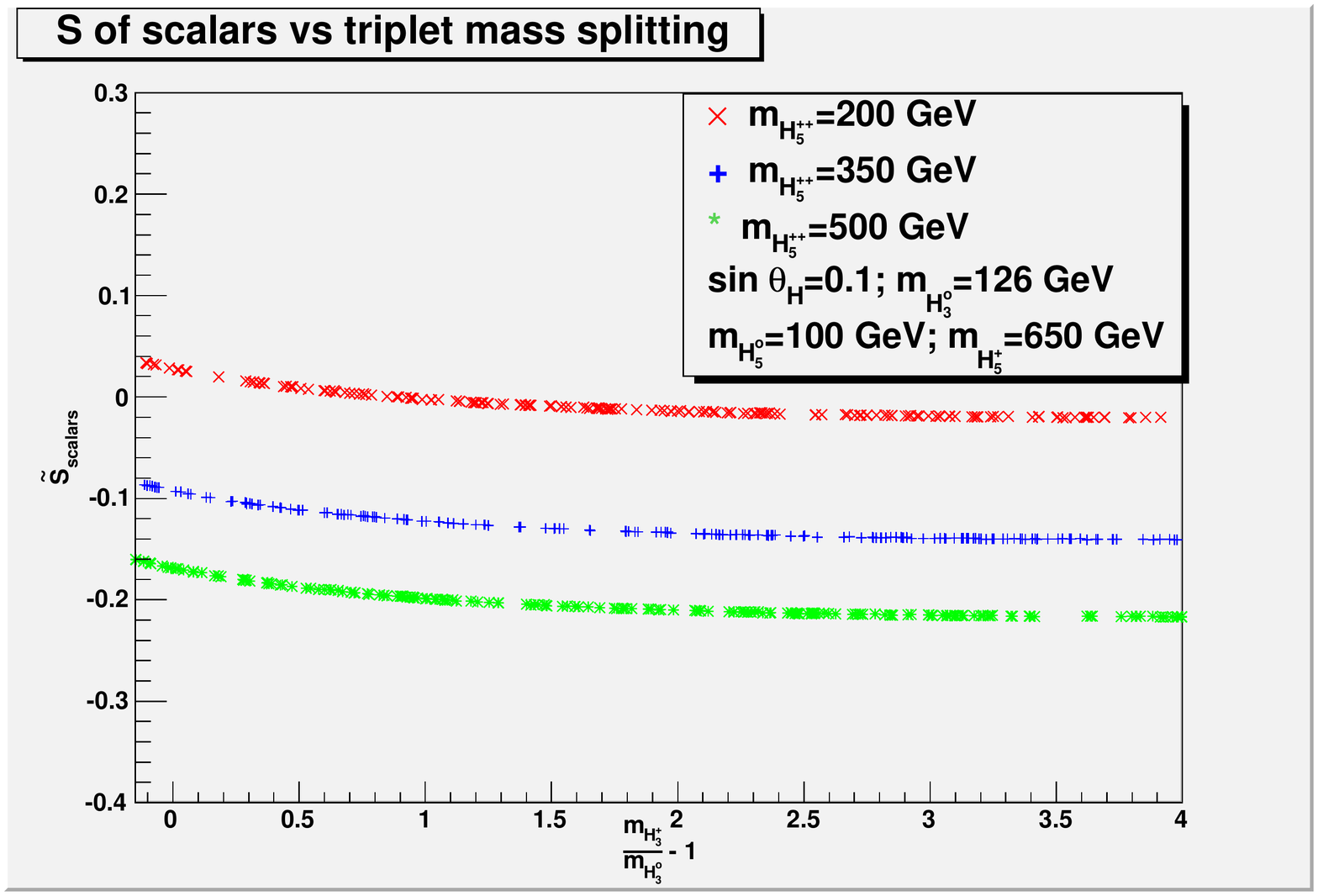} 
    \vspace{-2pt}
  \includegraphics[scale=0.3]{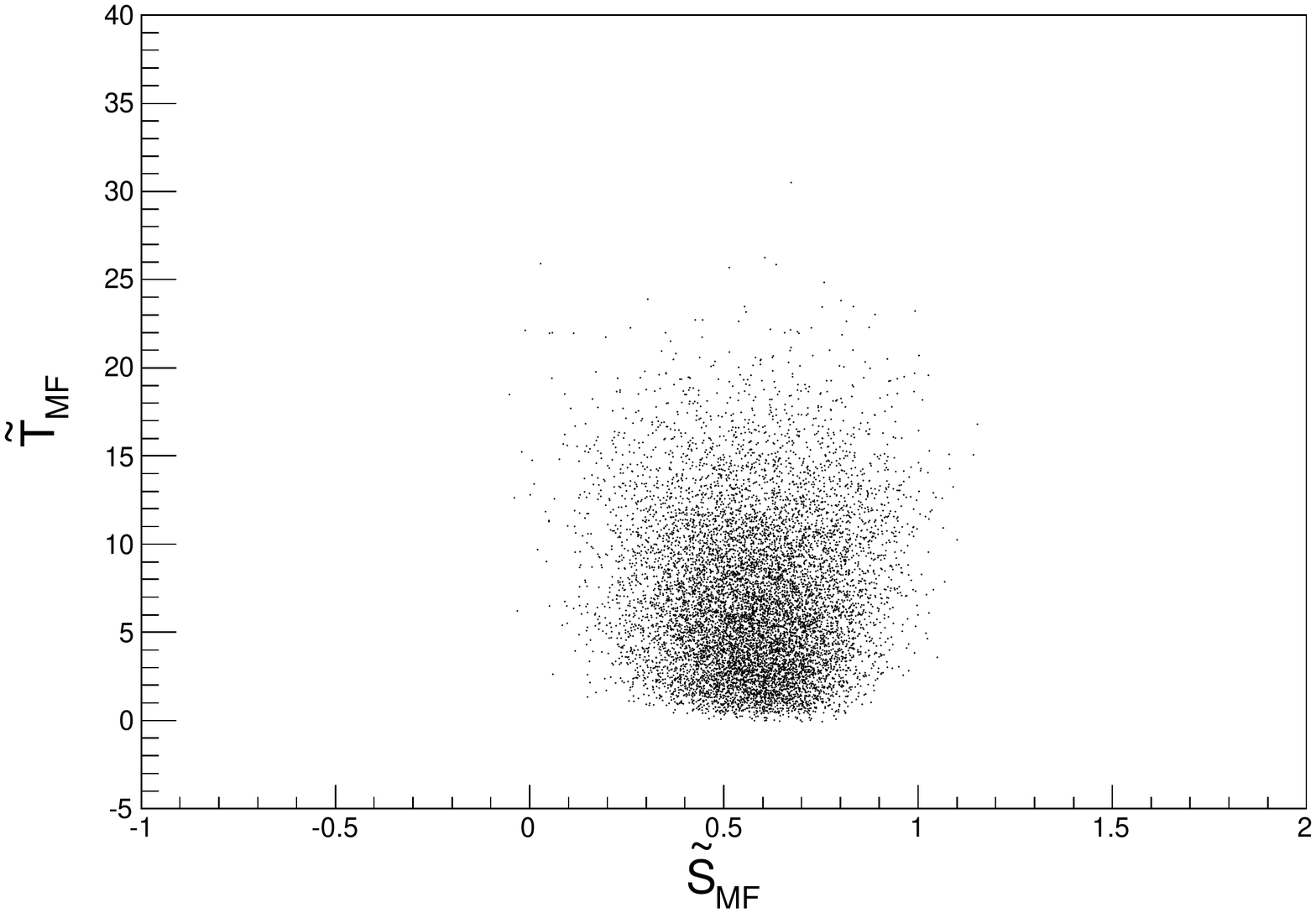}    
\caption{}
\label{Ss1}
\end{figure}

In the second plot of Fig.~\ref{Ss1}, $\tilde{S}{scalar} \equiv \tilde{S}_S$. The first plot in Fig.~\ref{Ss1} shows the contributions to T and S from the scalar sector for a large set of parameters with 10,000 combinations of mass and mixing parameters. As emphasized in \cite{ajinkya}, {\em those points (not an exhaustive list) are chosen for values of masses consistent with perturbative constraints}.This first scatter plot shows a wide range of values for S and T, out of which the region $\tilde{T}_S \approx 0$ and $\tilde{S}_S \approx 0$ occupies a small fraction of the parameter space. The next plot in Fig.~\ref{Ss1}, shows, as an example, the variation of $\tilde{S}_S$ as a function of the mass splitting within the $SU(2)_D$ triplet for a set of mass and mixing parameters.

Although one can always find a combination of parameters for which $\tilde{S}_S \approx 0$, the above plots show that there is a {\em large parameter space} for which $\tilde{S}_S \neq 0$ where there could exist a wide range of mass splittings among the scalars. If we could ascribe probabilities to the points in the first plot of Fig.~\ref{Ss1}, it is not unreasonable to say that the parameter region where $\tilde{T}_S \approx 0$ and $\tilde{S}_S \approx 0$ has a "{\em low probability}" and hence is more subject to "{\em fine tuning}". A large fraction of the parameter space would be excluded by the present constraint on $S$ if there are no additional degrees of freedom in the model.

\begin{figure}[H]
\centering
    \includegraphics[scale=0.3]{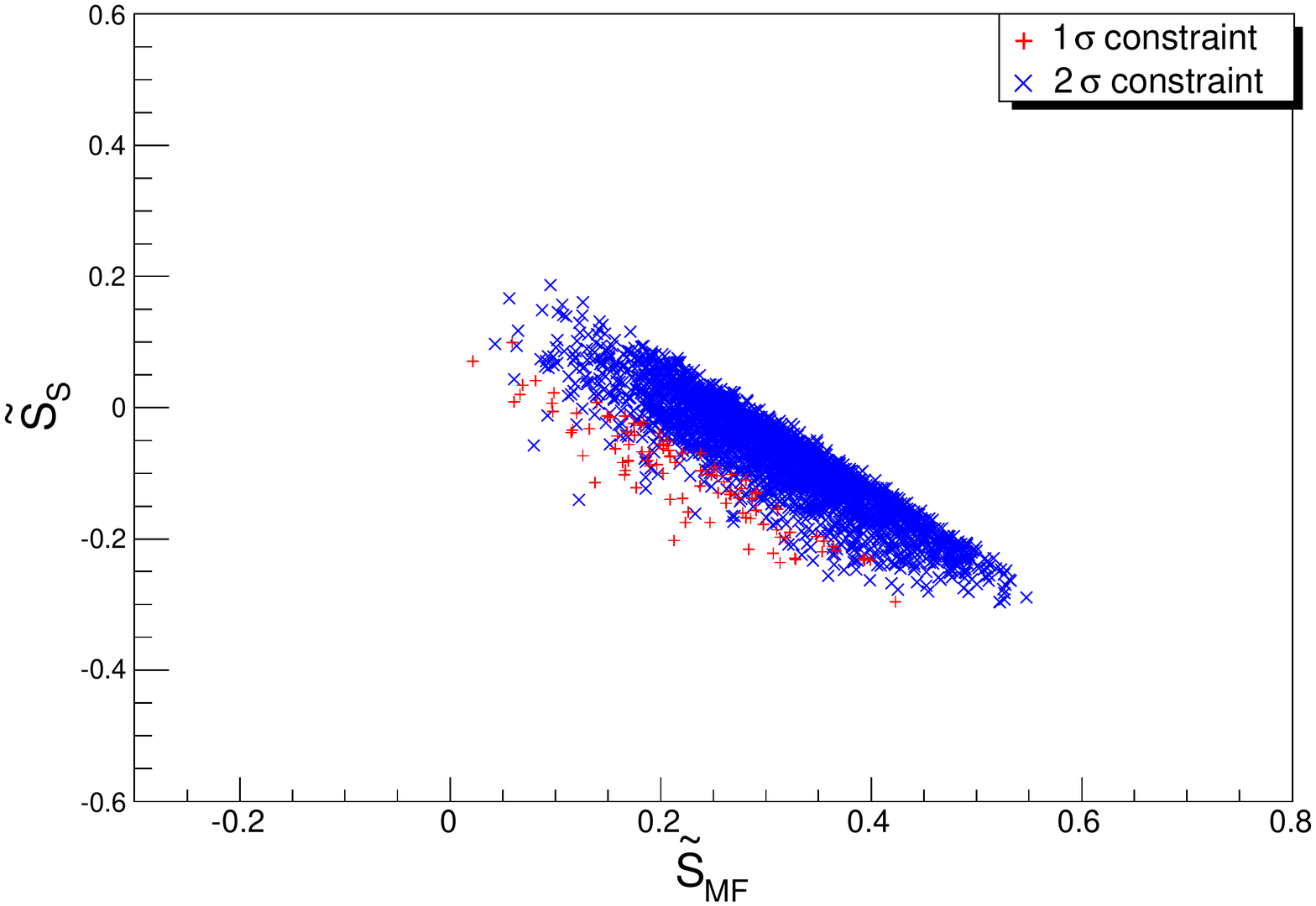} 
\vspace{-3pt}
    \includegraphics[scale=0.3]{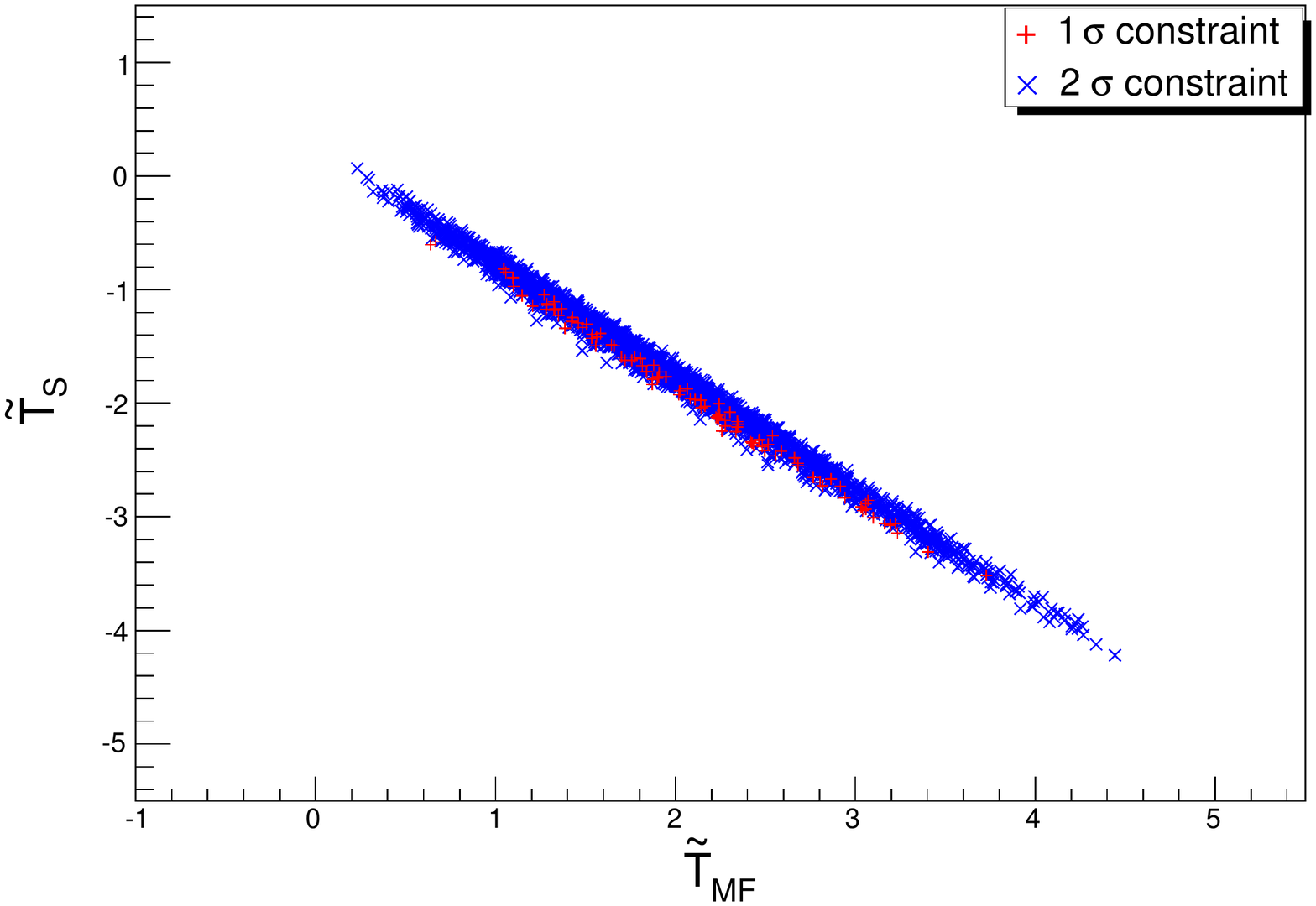} 
\caption{}
\label{TsTmf}
\end{figure}

Fig.~\ref{TsTmf} shows what these additional degrees of freedom might be: the mirror fermions of the Electroweak-scale right-handed neutrino (EW-$\nu_R$) model \cite{pqnur}. 
As shown in \cite{ajinkya}, the new physics contributions to S and T are now $\tilde{S} = \tilde{S}_S + \tilde{S}_{MF}$ and $\tilde{T}=\tilde{T}_S + \tilde{T}_{MF}$, where $MF$ stands for mirror fermions. First, for the set of masses consistent with perturbative constraints, the third plot in Fig.~\ref{Ss1} shows a scatter plot of the contributions of mirror fermions to S and T. 
Second, as expected, the contributions of mirror fermions to the S-parameter are {\em positive} since they belong to $SU(2)$ chiral (right-handed) doublets. It is these positive contributions from mirror fermions which could potentially "cancel" the negative contributions from the scalar triplets as shown in Fig.~\ref{TsTmf}. 

In these plots, the red and blue points refer to the 1 $\sigma$ and 2 $\sigma$ constraints based on earlier values of S and T, namely $S=-0.02 \pm 0.14$ and $T=0.06 \pm 0.14$. Although the plots shown here will be updated with the new values $S=0.02 \pm 0.07$ and $T= 0.06 \pm 0.06$, it is sufficient to show the trend of cancellations between the scalar and mirror fermion sectors. 
\vspace{-7pt}
\setlength\belowcaptionskip{+7ex}
\section{Conclusions}
In this note, we have discussed implications of electroweak precision constraints, in particular the S-parameter, on models containing Higgs triplets. Presently, $S=0.02 \pm 0.07$ and $T= 0.06 \pm 0.06$ with projected errors to be further reduced from data obtained at planned electron-positron colliders. Based on the current and projected precision of the electroweak precision parameters, it is found that models which contain Higgs triplets in addition to the SM particle content are tightly constrained in terms of mass splitting within the scalar triplet, to the point of "fine tuning" as the precision improves in the future. It is also found that models which contain additional mirror fermions beside the scalar triplets, such as the EW-$\nu_R$ model \cite{pqnur}, can {\em evade such constraints} due to the cancellations between the scalar and mirror fermion sectors.
\vspace{-6pt}

\begin{acknowledgements}

I wish to thank Alfredo Aranda and Tc Yuan for comments. I also wish to thank Liantao Wang for information on future electroweak precision measurements at the FCC. Finally, I wish to thank my student Dat Duong for help in Fig.~\ref{S1}.
\end{acknowledgements}


\vspace{6.cm}

\begin{thebibliography}{50}
\bibitem{GM}
H.~Georgi and M.~Machacek,
  Nucl.\ Phys.\ B {\bf 262}, 463 (1985);
  M.~S.~Chanowitz and M.~Golden,
  Phys.\ Lett.\  {\bf 165B}, 105 (1985).
  \bibitem{pqnur}
   P.~Q.~Hung,
  Phys.\ Lett.\ B {\bf 649}, 275 (2007).
  \bibitem{models}
  G.~Senjanovic and R.~N.~Mohapatra,
  Phys.\ Rev.\ D {\bf 12}, 1502 (1975);
  R.~N.~Mohapatra, A.~Perez-Lorenzana and C.~A.~de Sousa Pires,
  Phys.\ Lett.\ B {\bf 474}, 355 (2000).
  \bibitem{mehrdad}
   M.~Adibzadeh and P.~Q.~Hung,
  Phys.\ Rev.\ D {\bf 76}, 085002 (2007)
  \bibitem{PDG}
  M. Tanabashi et al. (Particle Data Group), Phys. Rev. D 98, 030001 (2018).
  \bibitem{reece}
  J.~Fan, M.~Reece and L.~T.~Wang,
  JHEP {\bf 1509}, 196 (2015);
  M.~Reece,
  Int.\ J.\ Mod.\ Phys.\ A {\bf 31}, no. 33, 1644003 (2016).
  \bibitem{atlas}
  G.~Ucchielli [ATLAS Collaboration],
  PoS EPS {\bf -HEP2017}, 722 (2017).
  \bibitem{cms}
  CMS Collaboration [CMS Collaboration],
  CMS-PAS-HIG-16-036.
  \bibitem{ajinkya}
   V.~Hoang, P.~Q.~Hung and A.~S.~Kamat,
  Nucl.\ Phys.\ B {\bf 877}, 190 (2013).
 
  
\end{thebibliography}
\end{document}